# Fourier optics with linearly tapered waveguides: light trapping and focusing

Mahmoud A. Gaafar[1,2,*], Hagen Renner[1], Manfred Eich[1,3], and Alexander Yu. Petrov[1,3,4]

[1]*Institute of Optical and Electronic Materials, Hamburg University of Technology, Eissendorfer Strasse 38, 21073 Hamburg, Germany*
[2]*Department of Physics, Faculty of Science, Menoufia University, Menoufia, Egypt*
[3]*Institute of Materials Research, Helmholtz-Zentrum Geesthacht, Max-Planck-Strasse 1, Geesthacht, D-21502, Germany*
[4]*ITMO University, 49 Kronverkskii Ave., 197101, St. Petersburg, Russia*
[*]*Corresponding author:* mahmoud.gaafar@tuhh.de

**Abstract:** An optical pulse asymptotically reaching zero group velocity in tapered waveguides can ultimately stop at a certain position in the taper accompanied by a strong spatial compression. This phenomenon can be also observed in spatio-temporal systems where the pulse velocity asymptotically reaches the velocity of a tapered front. The first system is well known from tapered plasmonic waveguides where adiabatic nano-focusing of light is observed. Its counterpart in the spatio-temporal system is the optical push broom effect where a nonlinear front collects and compresses the signal. Here, we use the slowly-varying envelope approximation to describe such systems. We demonstrate an analytical solution for the linear taper and the piecewise linear dispersion and show that the solution in this case resembles that of an optical lens in paraxial approximation. In particular, the spatial distribution of the focused light represents the Fourier transform of the signal at the input.

## Introduction

Light can be stopped if it is transferred into the mode with zero group velocity. There are two possibilities how zero group velocity can be obtained in the dispersion relation. One is the case in the middle of the dispersion relation where the mode continues at larger and smaller wavenumbers with non-zero group velocity. This is the case of modes at photonic band edges [1],[2] and in specially engineered waveguides [3]–[5]. Another case relates to zero group velocity asymptotic at large wavenumbers, which is observed in plasmonic waveguides [6]. This asymptotic case is particularly appealing as in adiabatically tapered waveguides light can stop in the taper [7],[8] without reflection. In contrast, zero group velocity at finite wave numbers cannot lead to ultimate light stopping and light will leave the taper in the forward or backward direction [9]–[12]. The asymptotic zero group velocity can also be considered in a more general way when the taper or a front is moving along the waveguide. In this case the zero group velocity of the light pulse is measured in respect to the front. Thus, the dispersion relation of the signal should asymptotically approach at large wave numbers the linear slope with group velocity equal to the velocity of the front.

Light trapping/freezing is obtained in tapered metal–dielectric waveguides that support surface plasmon–polariton (SPP) modes. That occurs when a SPP encounters a tapered perturbation that shifts the dispersion curve vertically with propagation distance and does not allow further propagation at certain distance in the taper for a defined wavelength, leading to nanofocusing effects [13]–[15],[8] (Figs. 1(a) and (c)). Here, different frequency components of the light freeze at different positions in the taper, leading to the spatial separation of its spectrum, as shown in Fig. 1(c). This effect has been demonstrated using different nanofocusing structures, including metal [16] and dielectric wedges [17],[8].

Alternatively, moving perturbations can be considered in nonlinear optics. Optical pulse propagation in dispersive waveguides and its dynamic control via a nonlinearly generated refractive index front (free carrier or Kerr effect) in the same waveguide has caught the attention in recent years [18]–[23]. In particular, so called "optical pushbroom effect" is discussed [24]–[26], in which the signal trapping is induced by a refractive index front moving with a constant velocity in a waveguide with hyperbolic dispersion (Fig. 1(b)). The velocity of the front is equal to the asymptotic velocity of the hyperbolic dispersion relation. In this case, the slow light signal is trapped inside a fast co-propagating front. In other words, the initial slowly propagating signal –after interacting with the front- is accelerated up to the front velocity and further interaction does not lead to group velocity change, as illustrated schematically in Fig 1(d). This effect has been theoretically proposed by de Sterke [26] and experimentally realized in a fiber Bragg grating [24] and in silicon Bragg grating recently [25]. Such trapping leads to the pulse compression as the energy of the input signal is concentrated inside the front.

The description of tapered plasmonic waveguides was based up to now on ray optics approximation [27],[8]. The optical push broom description was also based on ray optics approach [18] or coupled wave equations [28], the latter applicable only for waveguides with weak periodic perturbation. Here we use the slowly-varying envelope approximation with temporal evolution of spatial profiles to describe such tapers [29]. For a linear taper and piecewise linear dispersion curve an analytical solution is derived that shows that such a taper has strong similarity with conventional lenses. Namely, the spatial profile of the trapped light inside that taper represents the Fourier transform of the spatial profile at the input. In case of a moving perturbation the same applies to the temporal profile, that leads to a new realisation of the so-called time lens effects [30],[31]. In analogy to the spatially quadratic phase shift introduced by spatial lenses, a front induced-time lens imparts a temporally quadratic phase shift to an input signal wave. With that we show that tapered waveguides with dispersion curves asymptotically approaching constant velocity can be used for Fourier optics, including signal focusing and conversion between spatial frequencies and spatial profiles.

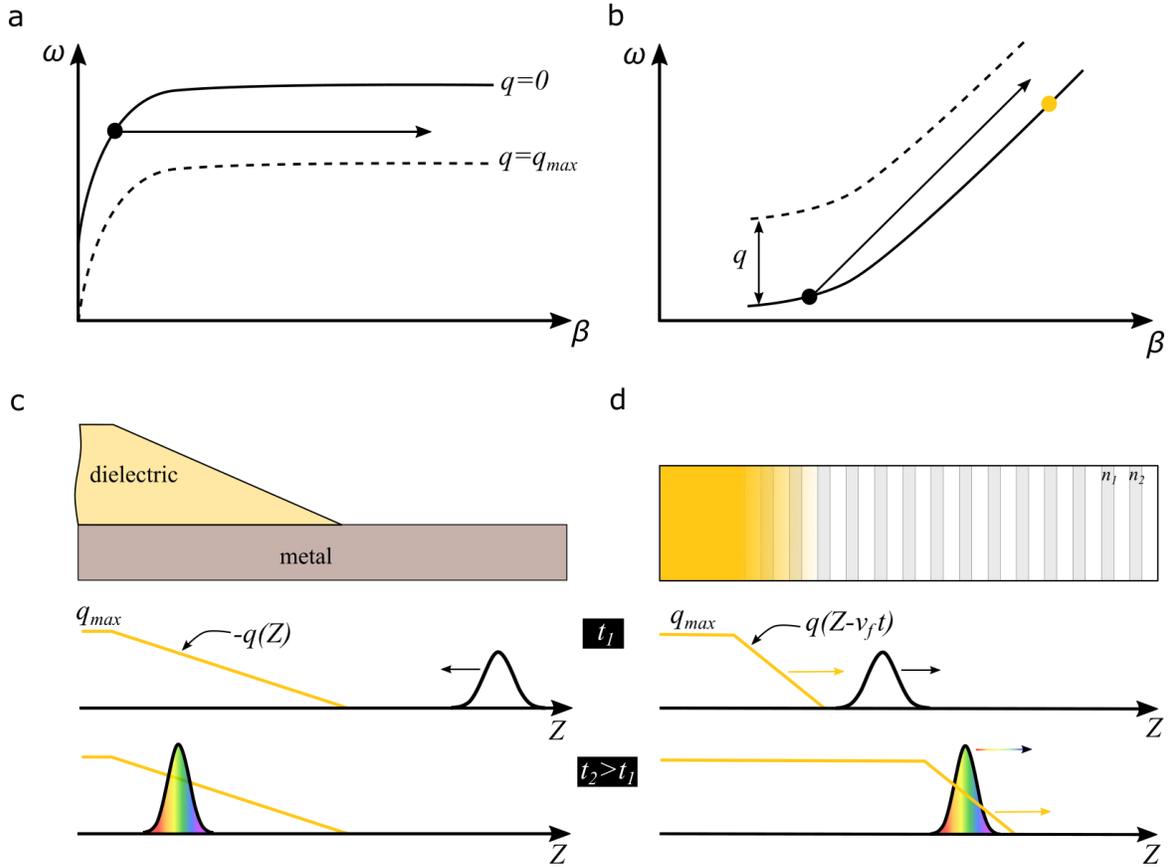

Fig. 1: Schematic illustration of light trapping in tapered plasmonic waveguides a and c and by optical push broom effect in nonlinear fronts b and d. (a) and (b) are the schematic dispersion relations corresponding to the structures shown in (c) and (d), respectively. The solid curves represent the dispersion relations of the original waveguides, while the dashed curves represent the maximal frequency-shifted dispersion relations at the end of the taper or behind the front. Black dots represent the locations of the input signal on the band diagram, while orange circle in (b) represents the location of the pulse generating the front. In case of a tapered plasmonic waveguide, different frequency components of a guided wave packet stop at correspondingly different depths inside the taper, leading to the spatial separation of its spectrum [Fig. (c) bottom]. The same is obtained with a moving front $q(Z - v_f t)$ in a waveguide with a hyperbolic dispersion. Different spectral components will stop at slightly different locations z inside the front.

**Theory**

In order to analyze the interaction of an optical signal with a taper we employ a slowly varying envelope approximation resulting in the time-evolution Schrödinger equation [29]:

$$\frac{\partial a(t,z)}{\partial t} = (v_\text{f} - v_\text{g})\frac{\partial a(t,z)}{\partial z} + \sum_{m=2}^{\infty} i^{m+1}\frac{\omega_m}{m!}\frac{\partial^m a(t,z)}{\partial z^m} + iq(z)a(t,z) \qquad (1)$$

where $t$ and $z = Z - v_\text{f}t$ are the time and the retarded longitudinal propagation distance, respectively, $Z$ is the laboratory frame longitudinal coordinate and $v_\text{f}$ is the velocity at which the perturbation propagates. The slowly varying amplitude of the signal is denoted as $a(t,z)$, and $q(z) = q(Z - v_\text{f}t)$. This approximation describes systems where the dispersion relation is shifted vertically in frequency by the perturbation without significant change of its shape [29].

The taper is moving in the $(t,Z)$-coordinates (laboratory frame) and seemingly standing in the $(t,z)$-coordinates. Here it has been assumed that in the absence of the perturbation $[q(z) = 0]$ individual waves of spatial frequency $\beta$ propagate in proportionality to $\exp[i(\omega t - \beta Z)]$ into the positive $Z$-direction and that the optical angular frequency $\omega$ is related to a given spatial frequency $\beta$ by the dispersion relation $\omega(\beta)$. The coefficients of the spatial derivatives in the sum result from the Taylor expansion of the dispersion relation as the $m$-th derivatives $\omega_m = \text{d}^m\omega/\text{d}\kappa^m|_{\kappa=0}$ at a certain reference spatial frequency $\beta_0$ of the Fourier spectrum of the signal, where $\kappa = \beta - \beta_0$. The group velocity $v_\text{g}$ is equivalent with $\omega_1$ and we assume $v_\text{g} < v_\text{f}$ throughout this work. In the laboratory frame, the total signal field thus propagates as

$$a(t, Z - v_\text{f}t)\exp[i(\omega_0 t - \beta_0 Z)] \qquad (2)$$

and $a(t, Z - v_\text{f}t) = a(t,z)$ varies much more slowly with time than $\exp(i\omega_0 t)$ does. The dispersion relation of the dispersive medium is now assumed to be piecewise linear,

$$\omega(\beta_0 + \kappa) = \begin{cases} \omega_0 + \omega_1\kappa & \text{for} \quad -\infty < \kappa \leq \kappa_\text{K}, \\ \Omega_0 + \Omega_1\kappa & \text{for} \quad \kappa_\text{K} \leq \kappa < \infty, \end{cases} \qquad (3)$$

where $\Omega_0 = \omega_0 + (\omega_1 - \Omega_1)\kappa_\text{K}$ guarantees continuity at $\kappa = \kappa_\text{K}$, see Fig. 2(b). The angular optical frequency at $\beta = \beta_0$ or $\kappa = 0$ is $\omega_0$. Such a piecewise linear function can be an approximation of the dispersion relation close to an anti-crossing between two interacting modes.

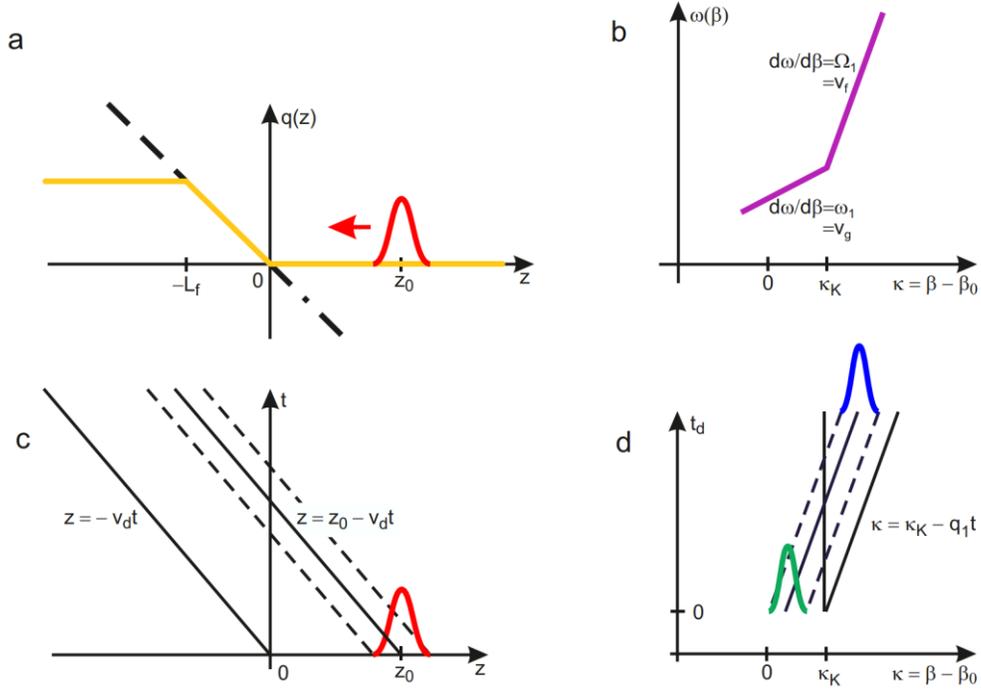

Fig. 2: (a) Relative movement of a signal (red line) into a moving index front (orange line) and assumed extensions of the front slope towards $z \to -\infty$ (black dashed line) and $z \to +\infty$ (black dash-dotted line). (b) Piecewise linear dispersion curve with a change of the slope at $\kappa = \kappa_K$. (c) Relative movement of the signal from $z = z_0$ into the front at $z \leq 0$. (d) Relative movement of the signal spectrum (green line) from the first branch of the dispersion curve ($\kappa \leq \kappa_K$) into the second branch ($\kappa \geq \kappa_K$). Blue line represents the output signal.

In the beginning at $t = 0$ the signal is assumed to be completely outside the taper [concentrated around $z = z_0$ in Fig. 2(a)], and its spatial spectrum be lumped around $\kappa = 0$ with a width much smaller than $\kappa_K$. Thus, the signal initially 'sees' an unperturbed medium and the first linear branch of the dispersion relation $\omega(\beta_0 + \kappa) = \omega_0 + \omega_1 \kappa$. With $\omega_m = 0$ for all $m \geq 2$, eq. (1) now simplifies to

$$\frac{\partial a(t,z)}{\partial t} = v_d \frac{\partial a(t,z)}{\partial z} + iq(z)a(t,z) \qquad (4)$$

where the differential velocity $v_d = v_f - v_g > 0$ was introduced. For positive $v_d$ the signal is moving to negative $z$. The solution for an initial signal $a(t = 0, z) = a_0(z)$ lumped around a position $z_0$ outside the taper can be written [32]

$$a(t,z) = a_0(z + v_d t) \exp\left[\frac{i}{v_d} \int_z^{z+v_d t} q(z') dz'\right] \qquad (5)$$

In this solution, a continuous linear shift of the position of the original envelope by $-v_d t$ [see Fig. 2(c)] and a phase modified in dependence of the front shape $q(z)$ can be observed. In particular, if a constant $q(z) = q_0$ is assumed the exponential function in eq. (5) becomes

$\exp(iq_0 t)$ indicating that $q_0$ acts, according to eq. (2), as a change in temporal angular frequency from $\omega(\beta)$ to $\omega(\beta) + q_0$, i.e., as a vertical shift of the dispersion curve.

We are interested in $q(z)$ which is not changing with time and is tapered along z. We specify it to have a linear slope section as

$$q(z) = \begin{cases} -q_1 L_f & \text{for} \quad -\infty < z \leq -L_f, \\ q_1 z & \text{for} \quad -L_f \leq z \leq 0, \\ 0 & \text{for} \quad 0 \leq z < \infty, \end{cases} \quad (6)$$

where $L_f$ is the length of the taper [Fig. 2(a)] and $q_1 \leq 0$. As will be shown later the signal will come to rest at a certain position inside the taper if the section of the constant slope and hence $L_f$ is long enough. Therefore, we can also assume $L_f \to \infty$ in the following [see dashed line in Fig. 2(a)]. Since in the beginning the signal was completely outside of the taper and afterwards the unchanged envelope moves at a relative velocity of $-v_d$ into the taper we are only interested in the solution inside the angle formed by the two lines $z = 0$ and $z = -v_d t$ [see Fig. 2(c)] for which the integral in eq. (5) reads $\int_z^{z+v_d t} q(z') dz' = -q_1 z^2 / 2$ for $-v_d t \leq z \leq 0$ and the solution is

$$a(t, z) = a_0(z + v_d t) \exp\left(-\frac{i q_1 z^2}{2 v_d}\right) \quad \text{for} \quad -v_d t \leq z \leq 0, \quad (7)$$

with the central position at $z = \hat{z}(t) = z_0 - v_d t$ which crosses the entrance of the front at time $t = z_0 / v_d$ [Fig. 2(c)]. Thus, we see that the transition through the front leads to the introduction of a spatially quadratic phase chirp on the signal. This is similar to what a lens is doing to the spatial profile of a beam, but in the transverse coordinates $x$ and $y$. It should be mentioned that if the onset of the linear slope is not sharp and has a smooth transition, instead, then still the same quadratic phase chirp will finally be obtained. All parts of the signal accumulate the same phase shift when they propagate though the smooth transition.

When the taper is moving as a front, $v_f \neq 0$, we can also use laboratory frame coordinates $(t, Z)$ in eq. (7)

$$a(t, Z - v_f t) = a_0(Z - v_g t) \exp[i \Psi(t, Z)] \quad \text{for} \quad +v_g t \leq Z \leq v_f t, \quad (8)$$

where the additional phase $\Psi(t, Z)$ caused by the front now depends on both coordinates

$$\Psi(t,Z) = -\frac{q_1(Z^2 - 2Zv_\mathrm{f}t + v_\mathrm{f}^2 t^2)}{2v_\mathrm{d}}. \qquad (9)$$

At the moving position $Z = \hat{Z}(t) = Z_0 + v_\mathrm{g}t = z_0 + v_\mathrm{g}t$ where the signal is lumped around we obtain the additional temporal angular frequency as

$$\Delta\omega[t,\hat{Z}(t)] = +\left.\frac{\partial \Psi(t,Z)}{\partial t}\right|_{Z=\hat{Z}(t)} = v_\mathrm{f} q_1 \left(\frac{Z_0}{v_\mathrm{d}} - t\right) \qquad (10)$$

while the additional spatial frequency is found to be

$$\Delta\beta[t,\hat{Z}(t)] = -\left.\frac{\partial \Psi(t,Z)}{\partial Z}\right|_{Z=\hat{Z}(t)} = q_1 \left(\frac{Z_0}{v_\mathrm{d}} - t\right) \qquad (11)$$

meaning that the shifts in temporal and spatial frequency are proportional to each other as $\Delta\omega[t,\hat{Z}(t)] = +v_\mathrm{f}\Delta\beta[t,\hat{Z}(t)]$. The two different signs in front of the derivatives in eqs. (10) and (11) have been chosen to make both increments add up positively to its corresponding frequency, $\omega_0$ and $\beta_0$, respectively, according to eq. (2).

In particular, eq. (11) indicates a continuous increase of the total spatial frequency by a value $-q_1 t$ linear increasing ($q_1 < 0$) with time while the front runs further and further through the signal. The accompanying increment in temporal frequency is $-q_1 v_\mathrm{f} t$. The increment in spatial frequency shifts the spectrum of the signal ever closer towards the kink of the dispersion curve at $\kappa = \kappa_\mathrm{K}$ [see Fig. 2(d)] where the assumed linearity of the dispersion curve and hence the solutions (5) and (7) are no longer valid.

In order to proceed with the analysis with this more general dispersion relation deviating from a purely linear one we may switch over to the spatial frequency domain. In doing so we may exploit the fact that the signal is now completely inside the slope of the front ($-L_\mathrm{f} \leq z \leq 0$) and approximate the latter as $q(z) = q_1 z$ in the infinite range $-\infty < z < +\infty$ [see dashed and dashed-dotted lines in Fig. 2(a)]. The Fourier transform of eq. (1) with Fourier kernel $\exp(+i\kappa z)$ is

$$\frac{\partial B(t,\kappa)}{\partial t} = i\overline{\omega}(\kappa)B(t,\kappa) + q_1 \frac{\partial B(t,\kappa)}{\partial \kappa} \qquad (12)$$

where $B(t,\kappa)$ is the Fourier transform of $a(t,z)$ and $\overline{\omega}(\kappa) = \omega(\beta_0 + \kappa) - \omega_0 - v_\mathrm{f}\kappa$ is the dispersion relation in the retarded coordinate system. This equation has the same principal form as eq. (4). Thus, with a starting spectrum $B_0(\kappa)$ at a time $t = t_0$ at which the real-space signal has already fully entered the front the solution is [32]

$$B(t, \kappa) = B_0(\kappa + q_1 t_d) \exp\left[-\frac{i}{q_1} \int_{\kappa+q_1 t_d}^{\kappa} \overline{\omega}(\kappa') d\kappa'\right] \tag{13}$$

where $t_d = t - t_0$. With the assumed $q_1 < 0$ the envelope of the initial spectrum is still permanently shifted towards higher spatial frequencies by $-q_1 t$ while the phase is modified in dependence of the dispersion relation. At $t = t_0$ the spectrum is assumed to be still in the first linear branch of the dispersion relation [Fig. 2(b) and (d)]. Then it shifts towards larger spatial frequencies $\kappa$ and beyond the kink at $\kappa = \kappa_K$. Finally, it will move inside the angle given by the two lines $\kappa = \kappa_K$ and $\kappa = \kappa_K - q_1 t$ [see Fig. 2(d)].

The special assumption in this work is that the slope $\Omega_1$ of the second branch of the dispersion relation and thus the group velocity in that spectral range be equal to the velocity $v_f$ at which the front propagates [see Fig. 2(b)]. Thus we can write

$$\overline{\omega}(\kappa) = \begin{cases} -v_d \kappa & \text{for} \quad -\infty < \kappa \leq \kappa_K, \\ -v_d \kappa_K & \text{for} \quad \kappa_K \leq \kappa < \infty, \end{cases} \tag{14}$$

For the wavenumbers above $\kappa_K$ this slope-corrected dispersion relation becomes constant and thus has zero group velocity. Our relevant final solution within the spectral range $\kappa_K < \kappa < \kappa_K - q_1 t$ will be

$$B(t_d, \kappa) = B_0(\kappa + q_1 t_d) \exp\left[-\frac{i}{q_1}(p_0 + p_1 \kappa + p_2 \kappa^2)\right] \tag{15}$$

where $p_0 = v_d(\kappa_K^2 + q_1^2 t_d^2)/2$, $p_1 = v_d(q_1 t_d - \kappa_K)$ and $p_2 = v_d/2$. The starting spectrum $B_0(\kappa)$ at $t_d = 0$ is just the Fourier transform of $a(t, z)$ at $t = t_0$, and the spectrum evolves as

$$B(t, \kappa) = \exp\left[-\frac{i}{q_1}(p_0 + p_1 \kappa + p_2 \kappa^2)\right] \times$$
$$\times \int_{z=-\infty}^{\infty} a_0(z + v_d t_0) \exp\left(-\frac{i q_1 z^2}{2 v_d}\right) \exp[i(\kappa + q_1 t_d) z] dz$$

$$\tag{16}$$

The phase in the exponent in front of the integral has a quadratic dependence on $\kappa$. Thus, the transition through the kink of dispersion relation leads to an accumulation of quadratic phase, or different wavenumbers accumulating a linear shift in space in respect to each other. Here we again see the analogy to the lens optics. A paraxial beam f(x, y) with spatial cross section in x propagates along the y axis and experiences a quadratic spatial dispersion as different spatial frequencies $\kappa_x$ of the beam experience different spatial shifts in x, which increase with propagation distance. Thus, for a conventional lens the dispersion accumulates with propagation distance y, and at the focal length the Fourier transform of the beam is obtained. In the case of the taper, the pulse a(t, z) changes its spatial profile in z with propagation time t. But the parameter $p_2$, responsible for quadratic dispersion, is a constant and not a function of propagation time. Interestingly the accumulated dispersion is exactly sufficient to provide the focusing effect and Fourier transform of the spatial profile. The same as with the taper function the smooth transition between two linear functions in the dispersion relation does not change the accumulated quadratic phase if the signal completely changes to the second linear branch of the dispersion relation.

Inverse Fourier transform, interchange of the order of integration over $\kappa$ and $z$, using a Gaussian integral (e.g., formula 3.323.2 of Ref. [33]) and some straightforward mathematical rearrangements provide the real-space solution

$$a(t,z) = \sqrt{\frac{q_1}{2\pi i v_d}} \exp[i\Phi(t,z)] V(z) \qquad (17)$$

$$\Phi(t,z) = \frac{q_1}{2v_d} z^2 + (q_1 t - \kappa_K)z - v_d \kappa_K t \qquad (18)$$

$$V(z) = \int_{\zeta=-\infty}^{\infty} a_0(\zeta) \exp(+iK\zeta) d\zeta \qquad (19)$$

with $K = \kappa_K - zq_1/v_d$. Here, $V(z)$ is just the Fourier transform of the input signal as a function of the spatial frequency $K$. A certain marked feature in the shape of $V(z)$ at a certain $K = \widetilde{K}$, or equivalently at $z = \tilde{z} = v_d(\kappa_K - \widetilde{K})/q_1$, is thus "standing" at this position $z = \tilde{z}$ without any movement relative to the front. The shape of the Fourier transform $V(z)$ does not change with time. In addition, the Fourier integral is multiplied by an exponential phase term $\exp[i\Phi(t,z)]$. It contains a contribution $(q_1 t - \kappa_K)z$ indicating a permanent growth of the spatial frequency by $-q_1 t$ as also observed in the intermediate steps of the derivation above. Thus, the signal never stops changing and continuously moves in the dispersion relation towards larger wavenumbers. The fact that the group velocity in the retarded coordinate system stays zero at larger wavenumbers allows to keep the signal stopped for infinitely long times. If the dispersion relation would change at some point, the signal would start moving again. In addition, there appears a constant quadratic chirp in $z$ with chirp parameter $q_1/(2v_d)$. In order to cancel this phase contribution, the signal should be equipped with an initial respective prechirp. In case of conventional thin lenses, the object is placed at the focal plane before the lens, thus accumulating spatial dispersion by propagation till the lens. In our case the signal can be prechirped by a dispersive element with accumulated quadratic phase equal to $-p_2\kappa^2/q_1$.

Thus, with $z = Z - v_f t$ we have arrived at an explicit expression for the output signal in terms of the laboratory frame space and time coordinates $Z$ and $t$, respectively. The temporal behavior at the end of the interaction time, e.g., at the output of a photonic crystal waveguide, can be obtained by setting $Z$ equal to the spatial position of the output. By inserting $z = Z - v_f t$ it can be seen that in laboratory frame coordinates the result of the Fourier integral $V(Z - v_f t)$ travels at the same velocity $v_f$ as the front does while it keeps its shape independently of time. This is different from a lens where, behind the focus, the beam diverges. The laboratory frame position of a marked feature at $z = \tilde{z}$ travels at the front velocity as $\tilde{Z}(t) = v_f t - v_d(\kappa_K - \tilde{K})/q_1$.

The effect of the additional phase $\Phi(t,z)$ can be seen after writing it in laboratory frame coordinates as

$$\Phi(t, Z - v_f t) = \frac{q_1}{2v_d} Z^2 - q_1 \frac{v_g}{v_d} tZ - \kappa_K(Z - v_f t) + t^2 q_1 v_f \frac{2v_g - v_f}{2v_d} \qquad (20)$$

The corresponding additional spatial frequency at the traveling position $\tilde{Z}(t) = v_f t - v_d(\kappa_K - \tilde{K})/q_1$ becomes

$$\Delta\beta[t, \tilde{Z}(t)] = -\frac{\partial \Phi(t, Z - v_f t)}{\partial Z}\bigg|_{Z=\tilde{Z}(t)} = -q_1 t + 2\kappa_K - \tilde{K} \qquad (21)$$

indicating again a growth of the spatial frequency by an amount $-q_1 t$ linearly increasing with time $t$. Analogously, the additional temporal frequency at the traveling position $\tilde{Z}(t)$ is

$$\Delta\omega[t, \tilde{Z}(t)] = +\frac{\partial \Phi(t,Z)}{\partial t}\bigg|_{Z=\tilde{Z}(t)} = -q_1 v_f t + v_g(2\kappa_K - \tilde{K}) \qquad (22)$$

It also continuously increases linearly with time $t$. Obviously, an increase by $-q_1 t$ in the spatial frequency in eq. (21) goes along with an increase by $-q_1 v_f t$ in the temporal frequency of eq. (22) indicating a shift in the dispersion diagram with the slope $v_f$.

To be more specific we may assume a chirped Gaussian initial signal at $t = 0$,

$$a_0(z) = a(0, z) = \hat{a}_0 \exp\left[-\frac{(z-z_0)^2(1+i\Gamma)}{w_0^2}\right] \quad (23)$$

with an input amplitude $1/e$-width of $w_{\text{in}} = w_0$ in space and a chirp parameter $\Gamma$. The absolute value of the resulting output signal follows from eqs. (17)-(19) as

$$|a(t,z)| = \sqrt{\frac{q_1}{2v_d(1+\Gamma^2)}} \, \hat{a}_0 w_0 \exp\left[-w_0^2 \left(\frac{q_1 z - \kappa_K v_d}{2v_d\sqrt{1+\Gamma^2}}\right)^2\right] \quad (24)$$

"standing" at the position $z_s = \kappa_K v_d / q_1 < 0$ inside the front, or in laboratory frame coordinates, moving with the front and with its peak at the position $Z = v_f t + \kappa_K v_d / q_1$. The output amplitude $1/e$-width is $w_{\text{out}} = 2v_d\sqrt{1+\Gamma^2}/(w_0|q_1|)$. Two conditions must be fulfilled in order for the output signal to be almost completely inside the front: First, the output width must be small enough according to $w_{\text{out}} \ll |z_s|$ or, equivalently, $\kappa_K \gg 2\sqrt{1+\Gamma^2}/w_0$, where the right-hand side of the last inequality is exactly the $1/e$ spectral width of the input signal. Thus, choosing the input signal such that there is no spectral overlap with the second branch of the dispersion curve [$\kappa \geq \kappa_K$, see Fig. 2(b)] already ensures that the output signal will be practically inside the front. Second, in order to have the slope of the front long enough to keep the output signal inside the region of the front slope its length $L_f$ must be chosen such that the condition $w_{\text{out}} \ll L_f - |z_s|$ or, equivalently, $2v_d\sqrt{1+\Gamma^2} \ll w_0(L_f|q_1| - \kappa_K v_d)$ is fulfilled.

The compression factor defined as the ratio of output width to input width along the $z$ and the $Z$ axis, respectively, becomes

$$\eta_z = \eta_Z = \frac{w_{\text{out}}}{w_{\text{in}}} = \frac{2v_d\sqrt{1+\Gamma^2}}{w_0^2|q_1|} \quad (25)$$

It is smaller than one for $|v_d/q_1| < w_0^2/(2\sqrt{1+\Gamma^2})$, in which case the output signal is compressed as compared to the input signal. Thus, large longitudinal compression factors can be obtained for large length of input signal, sharp front and small velocity difference between input signal and the front causing a slow dive of the signal into the front. Since the input signal initially propagates at the group velocity $v_g$ whereas the output signal travels at the velocity $v_f$ of the front the compression factor in time is $\eta_t = \eta_z v_g / v_f$.

## Simulations

To confirm the obtained analytical solution we solved eq. (1) numerically by the split-step Fourier method [34]. We consider a moving linear perturbation in an optical push broom configuration (eq. 6), where $q_1 = \Delta q_{Dmax}/L_f$ is the front slope, $L_f = 2.25$ mm is the spatial front width and $\Delta q_{Dmax} = 1.5$ THz is the maximum vertical band diagram shift in frequency, cf. dashed black curve in Fig. 1(b). In addition, we use a piecewise linear dispersion relation, cf. Fig. 2(b). The group velocities of the signal and the front are chosen to be $c/4$ and $c/2$, respectively.

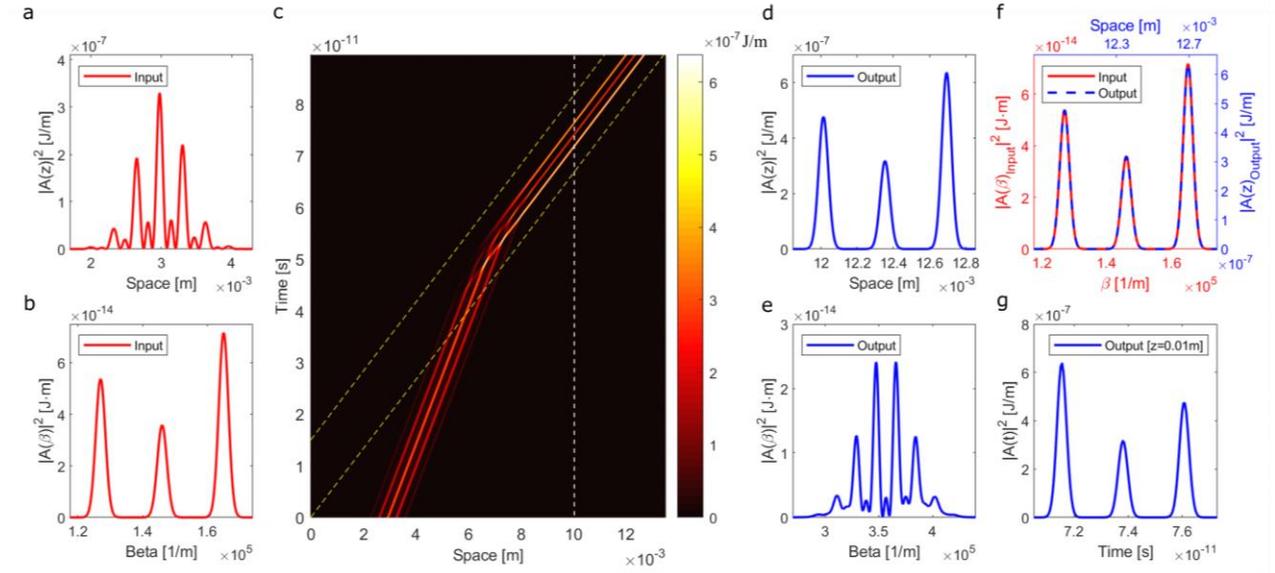

Fig 3: Simulation of pushbroom-induced time lens. (a) Spatial profile and (b) spectrum of 3 input Gaussian signal pulses with same duration (15 ps) and velocity ($c/4$) but with different amplitudes and center spatial frequencies launched at the same location in the waveguide. (c) Temporal evolution of the signal represented in the laboratory frame. The pseudo color indicates the energy density of the signal, while the dashed orange line marks the boundaries of the index front. The front velocity and duration are $c/2$ and 15 ps, respectively. Spatial profile and spectrum of the output signal pulse are presented in (d) and (e), respectively. (f) The input spectrum (solid red curve) and the output spatial profile (dashed blue curve) have been scaled and plotted together. (g) The temporal profile of the signal after it has been trapped inside the front at the location z=0.01 m (dashed white line in (c)).

Simulation results of pushbroom time lens are presented in Fig. 3. Here we launch three overlapping Gaussian signal pulses with same length (1.125 mm) but with different amplitudes and center spatial frequencies at the same location in the waveguide (Figs. 3(a) and (b)). Temporal evolution of the signal pulse represented in the stationary frame is shown in Fig. 3(c). The pseudo color indicates the power of the signal pulse. The dashed orange line marks the boundaries of the index front. As we can see, the trapped signal has spatial distribution frequency information (Fig. 3(b)) transferred into spatial information inside the front at the output (Fig.

3(d)). For ease of comparison, the input spectrum (solid red curve) and the output spatial profile (dashed blue curve) have been scaled and plotted together in Fig. 3(f), which proofs the Fourier transform property of pushbroom-induced trapping. The signal in the front also experiences the Fourier transform in time domain as well and with that the time lens effect (Fig. 3(g)).

In Fig. 4(a) we show the same results shown in Fig. 3 but represented in the frame moving with the front, which is equivalent to the case of a stationary taper in the waveguide. To validate our theory, we compare output spatial profiles and output spectra obtained from eq. (17) and numerical solution of linear Schrödinger equation (LSE) (eq. (1)) in Figs. 4(b)-(e). Indeed the analytical results fits very well the obtained results from numerical simulation.

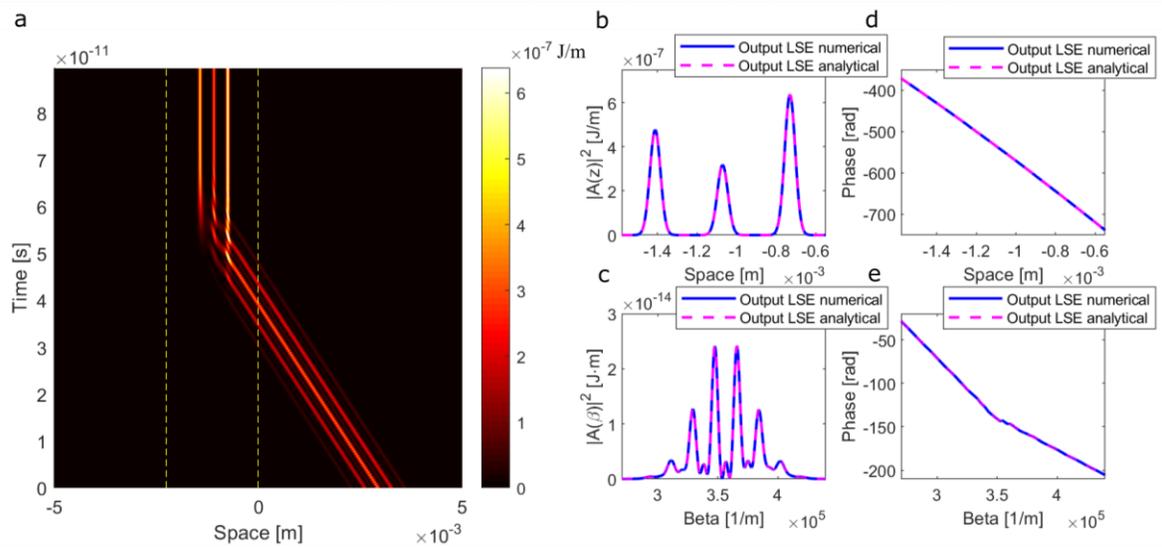

Fig 4: (a) Temporal evolution of the signal represented in the frame moving with the front. All simulation parameters are the same as in Figs. 3. (b)-(e) Comparison between the results obtained from eq. 17 of the analytical model (dashed magenta curves) and that obtained from numerical simulation using LSE (solid blue curves).

To demonstrate the pushbroom-induced lensing/compression effect, we simulate the front interaction with a wide Gaussian signal pulse (Fig. 5). The unchirped input signal pulse has a width of 3.75 mm (50 ps), while the front width is the same as before 2.25 mm (15 ps). As we can see in Fig. 5(a), the signal pulse after it has been completely approached by the index front is collected and compressed in space and time inside it, in analogy to the spatial compression induced by a normal lens. Compared to the case before, the wider input signal leads to a stronger spatial and temporal compression, as expected. The signal input/output spatial profiles and spectra are presented in Figs. 5(b) and (c), respectively. From eq. (25), and for $w_0$=1.59 mm,

$v_d = c/4$ and $|q_1| = 4.188$ THz/mm, this yields a compression factor of $\sim 7.06 \cdot 10^{-3}$ which is again confirmed by simulation results.

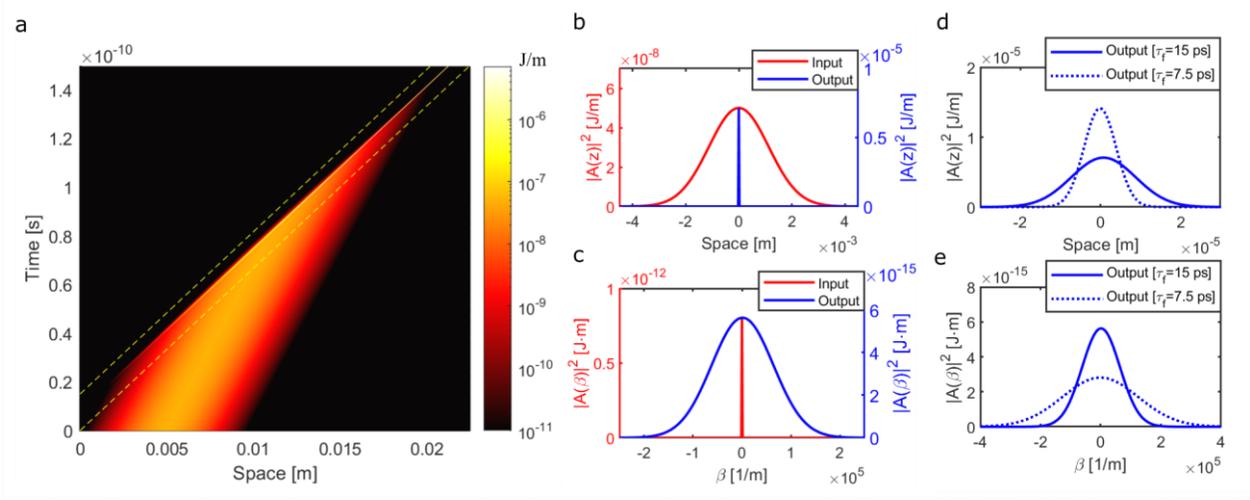

Fig 5: (a) Numerical study of the temporal evolution of a wide input Gaussian signal pulse represented in the stationary frame. The velocity and duration of the input signal pulse are $c/4$ and $\tau_s = 50$ ps, respectively, while the front velocity and duration are $c/2$ and 15 ps, respectively. (b) The spatial profiles and (c) the spectra of the input/output signal pulses. For ease of comparison we plot signal inputs and outputs shifted to zero. (d) Spatial profiles and (e) spectra of the output signal pulse in case of 15 ps-long (solid blue curves) and 7.5 ps-long (dashed blue curves) fronts.

We further investigate the effect of front's slope on the signal compression factor. For that, we used 2 times sharper front, while keeping other parameters the same as used for the simulation shown in Figs. 5(a)-(c). The comparison of the signal output spatial profiles and spectra using 15 ps (solid blue lines) and 7.5 ps (dotted blue lines) fronts are presented in Figs. 5(d) and (e). As can be seen, the compression factor increases 2 times when 2 times sharper front is used, as expected from eq. (25) too.

**Conclusion**

We have presented an analytical solution for the signal stopped in the tapered section of a waveguide where the velocity of the signal is approaching zero at large wavenumbers. This analytical solution shows strong analogy to lenses and corresponding Fourier optics. Namely, the stopped light has a spatial profile that is equivalent to the Fourier transform of the input signal spatial profile with a scaling factor that depends on the taper slope and waveguide dispersion. For the derivation, a piecewise linear dispersion and a linear front were considered. A smooth transition between two linear dispersion curves and smooth starting of the linear taper will lead to the same result. Thus, the results are also approximately describing the trapping in real tapers.

The provided solution applies to two systems from seemingly independent fields: nanofocusing in tapered plasmonic waveguides and optical push broom effect in nonlinear fronts. In both cases the input signal freezes inside the taper. Though the signal envelope stops in the taper, the spatial frequency continuously increase with time. In the case of plasmonic waveguides that will lead to stronger penetration of electric fields into the plasmonic medium and thus additional optical loss. However, for push broom trapping inside the front the losses are typically not a function of wavenumber and thus large compressions and trapping times can be envisaged.

The presented results pave the way for Fourier optics with linear tapers. Namely, linear tapers in plasmonic waveguides can be used to make spectral reconstructions as observed in rainbow trapping effects [35],[10]. The reverse effect can be used to structure the signal trapped in the taper. A frequency mix can be used to obtain light intensity with prescribed spatial profile in plasmonic tapers or pulses of defined form in moving fronts. For example, with push broom effect short pulses of defined temporal profile can be generated. The ability of pulse compression can be used to obtain nanofocusing in plasmonic tapes or large pulse compression in fronts.